# Isotopic evidence of long-lived volcanism on Io

Katherine de Kleer[1]*, Ery C. Hughes[1,2], Francis Nimmo[3], John Eiler[1], Amy E. Hofmann[4], Statia Luszcz-Cook[5,6,7], Kathy Mandt[8].

[1]Division of Geological and Planetary Sciences, California Institute of Technology, Pasadena 91125, USA.

[2]Earth Structure and Processes, Te Pū Ao | GNS Science, Avalon 5011, Aotearoa New Zealand.

[3]Department of Earth and Planetary Sciences, University of California Santa Cruz, Santa Cruz 95064, USA.

[4]Jet Propulsion Laboratory, California Institute of Technology, Pasadena 91109, USA.

[5]Liberal Studies, New York University, New York 10023, USA.

[6]Columbia Astrophysics Laboratory, Columbia University, New York 10027, USA.

[7]Department of Astrophysics, American Museum of Natural History, New York 10024, USA.

[8]NASA Goddard Space Flight Center, Greenbelt 20771, USA.

*Corresponding author. Email: dekleer@caltech.edu.

**Jupiter's moon Io hosts extensive volcanism driven by tidal heating. The isotopic composition of Io's inventory of volatile elements, including sulfur and chlorine, reflects its outgassing and mass loss history and provides an avenue for exploring its evolution. We used millimeter observations of Io's atmosphere to measure sulfur isotopes in gaseous $SO_2$ and SO, and chlorine isotopes in gaseous NaCl and KCl. We find $^{34}S/^{32}S=0.0595\pm0.0038$ ($\delta^{34}S=+347\pm86‰$), which is highly enriched compared to average Solar System values and indicates that Io has lost 94 to 99% of its available sulfur. Our measurement of $^{37}Cl/^{35}Cl=0.403\pm0.028$ ($\delta^{37}Cl=+263\pm88‰$) shows chlorine is similarly enriched. These measurements indicate that Io has been volcanically active for most or all of its history, with potentially higher outgassing and mass-loss rates at earlier times.**

Widespread volcanic activity on Jupiter's moon Io is powered by tidal heating of its interior, due to its orbital resonance with neighboring moons Europa and Ganymede. Models of the formation of Jupiter's large moons show that Io, Europa, and Ganymede were probably captured into the resonance during their formation process (*1, 2*). If this is the case, then Io and Europa have experienced strong tidal heating for the entire 4.57 Gyr history of the Solar System, implying that Io has been volcanically active (either continuously or cyclically) over the same time period (*3*).

Io's current volcanic activity resurfaces the moon at a rate of 0.1 to 1.0 cm yr$^{-1}$ (*4*). This has erased all impact craters from its surface (*5*), leaving a geological record of only the most recent million years of its history. However, isotopic abundances could record the history of volcanism on Io: if Io's current rates of mass loss [1000 to 3000 kg s$^{-1}$ (*6*)] and outgassing from its interior to its atmosphere have been sustained for billions of years, its reservoirs of volatile elements should be highly enriched in heavy stable isotopes, because atmospheric escape processes generally favor loss of lighter isotopes. Stable isotope measurements of volatile elements, such as sulfur and chlorine, could provide information on the history of volcanism at Io.

**Millimeter observations of Io**

We used the Atacama Large Millimeter/submillimeter Array (ALMA) to observe gases in Io's atmosphere (*7*). Io is tidally locked to Jupiter; the hemispheres facing into and away from its direction of motion, referred to as its leading and trailing hemispheres, were observed on 2022 May 24 and 2022 May 18 respectively (Universal Time, UT). We used ALMA's Band 8 receivers to cover the frequency range of 416 to 432 GHz (around 0.7 mm wavelength) in 13 spectral windows. This frequency range was chosen to cover multiple rotational transitions of SO$_2$, SO, NaCl, KCl, and their isotopologues, with the goal of determining the $^{34}$S/$^{32}$S and $^{37}$Cl/$^{35}$Cl ratios. The data were processed (*7*) to produce a calibrated spectral data cube with dimensions of right ascension (RA), declination (Dec), and frequency, with Io's thermal emission continuum subtracted. The data have a spectral resolution of 244 kHz (170 m s$^{-1}$) and an angular resolution of ~0.28 arcseconds (″) (equivalent to a spatial resolution of ~1000 km at the distance of Io at the time of our observations). The data cubes and extracted images for each species have a pixel scale of 0″.03, such that the spatial resolution is sampled with ~10 pixels per resolution element (*7*).

Figure 1 shows the spatial distribution of each of the four gas species we targeted. SO$_2$ and SO are concentrated in the low-to-mid latitudes, with the strongest emission close to the limb, where atmospheric path length is longest. In contrast, NaCl and KCl are confined to a few localized points, which we interpret as volcanic plumes (locations listed in Table S4).

We extracted disk-integrated spectra around the targeted emission lines from the data cubes (Figs. 2 & 3), using an aperture defined by all pixels in which the continuum emission is at least 5% of the peak continuum emission; this produces an aperture radius that is roughly 1.3× Io's radius. We quantify the noise in the spectrum for each of the 13 spectral windows independently, calculating it as the standard deviation of the disk-integrated spectrum in line-free regions.

**Atmospheric modeling**

We used a radiative transfer model of Io's tenuous atmosphere (*7, 8, 9*) to determine the $^{34}$S/$^{32}$S ratio, by fitting the observed emission lines of SO$_2$. SO$_2$ makes up ~90 to 97% of Io's atmosphere (*10*), so the abundances of all other species in the model are calculated as mixing ratios relative to $SO_2$. To determine the $^{34}S/^{32}S$ ratio, we jointly fitted two lines of $^{32}SO_2$ and four lines of $^{34}SO_2$. The SO lines were not used in the model fitting due to the low signal to noise ratio (SNR) of the $^{34}SO$ lines. The free parameters in the model are the $SO_2$ column density, the $^{34}S/^{32}S$ ratio, and the gas temperature. The gas temperature is constrained because the dataset includes both high- and low-excitation lines. The $^{34}SO_2$ and $^{32}SO_2$ lines were chosen to be sensitive to the same altitudes (*7*). As a result, the derived isotopic ratio is not strongly sensitive to the assumed atmospheric temperature profile.

The $^{37}Cl/^{35}Cl$ ratio was determined using the same procedure, by jointly fitting the observed NaCl and KCl lines for both chlorine isotopes (five lines total). The free parameters are the NaCl and KCl column densities (relative to $SO_2$), the gas temperature, the gas fractional surface coverage (the ratio of the emitting area of the gas to the projected surface area of Io), the $^{37}Cl/^{35}Cl$ ratio, and the line-of-sight gas velocity (relative to Io's velocity and rotation). Gas velocity is a free parameter in the chlorine model because the NaCl and KCl lines are all frequency-shifted from Io's rest frame, with each species shifted in the same way within each observation, indicating bulk motion of the chlorine-bearing gas relative to Io's rotation. This parameter was not necessary for fitting the $SO_2$ lines.

The observed lines are compared to the best-fitting models in Figures 2 and 3; the model parameters are listed in Table 1. The SO data are also shown in Figure 2 where they are compared to the model derived from $SO_2$; we find the SO data are consistent with the same isotope ratio as derived from $SO_2$. On Io, SO is produced by photo-dissociation of $SO_2$ (*11*), so we expect its $^{34}S/^{32}S$ ratio to only differ from that of $SO_2$ by a factor of 1.009 (*12*), which is smaller than the uncertainties. The best-fitting $SO/SO_2$ ratios are 3% and 5% for the leading and trailing hemispheres respectively, within the range of 3 to 10% found by previous studies (*10*, *13*).

The best-fitting gas temperatures for NaCl and KCl (assumed to be identical) are 774±66 and 682±56 K for the leading and trailing hemispheres respectively (all uncertainties are 1σ). These temperatures are consistent with a volcanic plume origin of NaCl and KCl, and within the previously reported range of 500 to 1000 K (*14*). In contrast, the leading and trailing hemisphere temperatures of $SO_2$, which is present primarily in Io's bulk atmosphere, are 225.9±3.3 and 240.1±7.5 K.

## The $^{34}S/^{32}S$ ratio

The leading and trailing hemispheres have $^{34}SO_2/^{32}SO_2$ ratios of 0.0543±0.0022 and 0.0646±0.0053 respectively, a difference of <1.5σ. Combining these, we find a global $^{34}SO_2/^{32}SO_2$ ratio of 0.0595±0.0038. This is within 2σ of the bottom of the previously reported range of 0.065 to 0.120 (*15*). The previous measurement used $^{32}SO_2$ and $^{34}SO_2$ lines that were sensitive to different altitudes (*7*), so was affected by degeneracies with the temperature profile.

We convert the derived isotope ratio to a $\delta^{34}S_{VCDT}$ value, defined as the difference between the measured value and the Vienna Canyon Diablo Troilite (VCDT) isotopic standard, which has $^{34}S/^{32}S = 0.04416$ (*16*). We find $\delta^{34}S_{VCDT} = +347±86‰$ for $SO_2$ in Io's atmosphere. Figure 4A compares this measurement to other Solar System bodies; we find that Io's atmosphere is more enriched in $^{34}S$ than most of these materials, by a large margin. The only measurement that reaches similarly high enrichment is $H_2S$ in Comet Hale-Bopp, but that measurement is still consistent with 0‰ within 2σ given its large uncertainties (Table S5).

We interpret the measured sulfur isotopic fractionation as due to a distillation process, whereby the lighter isotope is preferentially lost from a sulfur reservoir that is being continuously recycled between Io's interior and atmosphere. Atmospheric escape then distills the portion of Io's planetary inventory of sulfur that is available for recycling and loss.

Io's mass loss is driven by ion-neutral collisions between molecules in Io's atmosphere and energetic particles from the plasma in Jupiter's magnetosphere (*17*), rather than by thermal escape (*18*). Io's atmosphere is thought to be well-mixed up to a homopause [altitude ~30 km (*19*)]; between the homopause and exobase (~600 km), molecular diffusion produces a gravitationally stratified atmosphere (*20*), while the exosphere (above the exobase) is collision-less. Io's mass loss primarily occurs above the exobase because the conductive ionosphere diverts incoming plasma away from the near-surface region (*21*). From the homopause to the exobase, the partial pressure of each species decreases by $e^{-mg_z \Delta z/kT_z}$, where $m$ is the mass of the species (in kg mol$^{-1}$), $z$ is altitude (m), $g_z$ is Io's gravity (1.8 m s$^{-2}$ at the surface), k is the Boltzmann constant, and $T_z$ is the gas temperature at that altitude (K).

To determine how isotopically fractionated the material lost from Io's exosphere is, we compare the isotope ratio in SO$_2$ at the exobase to that at the homopause; for the latter we assume the isotopic ratio matches that of the well-mixed lower atmosphere. Adopting a temperature profile from previous work (*19*), we calculate that the material lost from Io's atmosphere has a $^{34}$S/$^{32}$S ratio 0.917 times that of the bulk atmosphere, which is the loss fractionation factor $^{34}\alpha_{loss}$ (see Supplementary Text).

If the atmosphere were in steady-state, with mass loss balanced by new material fed from a reservoir with $\delta^{34}$S$_{VCDT}$ ~0‰ (so with no distillation process), the atmospheric $\delta^{34}$S$_{VCDT}$ value would be 1-$^{34}\alpha_{loss}$ = +83‰. This steady state value is much lower than the +347±86‰ we measured, so we reject the possibility of a steady-state system without distillation.

A distillation process, consisting of recycling between the interior and atmosphere combined with mass loss from a gravitationally-stratified atmosphere, is consistent with additional constraints on the interactions between Io's surface and interior. If Io has maintained its current 0.1 to 1.0 cm yr$^{-1}$ resurfacing rate over the entire 4.57 Gyr history of the Solar System, and if that resurfacing predominantly occurs through volcanic deposition, then the volume of material required is 10 to 100 times Io's total volume. It is likely that Io's mantle participates in this cycle: Io's magmas are thought to be mantle material melted by tides and advected to the surface via heat pipes (*22, 23*). Therefore some fraction of the mantle, including its volatile elements, must have been recycled through the surface environment at least tens to hundreds of times.

**Io's mass loss history**

Assuming that loss from a gravitationally stratified atmosphere is the dominant isotopic fractionation process acting on Io's sulfur inventory, our calculated $^{34}\alpha_{loss}$ relates Io's sulfur isotope ratio today to the fraction of Io's initial sulfur that remains (*f*) via the Rayleigh equation for a constant fractionation factor $\alpha$:

$$^{34}R = {}^{34}R_0 f^{{}^{34}\alpha_{loss}-1} \tag{1}$$

where $^{34}R={}^{34}$S/$^{32}$S, and $^{34}R_0$ is its initial value in bulk Io (see Supplementary Text). If Io started with $^{34}R_0$ close to the Solar System average ($\delta^{34}$S$_{VCDT}$ ~0‰, Fig. 4A), our measured isotope ratio corresponds to $f = 0.028^{+0.033}_{-0.015}$, i.e. Io has lost 94 to 99% of its sulfur inventory that participates in the outgassing and recycling process.

In calculating $^{34}\alpha_{loss}$ above, we assumed that loss occurs only above the exobase and that gravitational stratification of the atmosphere is in the steady state. If atmospheric loss also occurs from altitudes below the exobase, or if Io's highly variable atmosphere does not reach a gravitationally stratified steady state, the loss process would produce less isotopic fractionation (higher $^{34}\alpha_{loss}$), so an even greater sulfur fraction would need to have been lost.

We next consider whether the current mass loss rate, acting on Io's initial sulfur inventory and sustained for 4.57 Gyr, would produce a loss fraction consistent with our measurement. Io is thought to have formed with a bulk composition close to that of ordinary chondrite meteorites classified as L or LL (24) which are ~2% S by mass (25). Adopting that composition and Io's current mass gives an estimate of $2\times10^{21}$ kg for Io's initial S mass. At Io's current mass loss rate of 1000 to 3000 kg s$^{-1}$ (6), and assuming that all sulfur loss is via SO$_2$, Io would have lost (1 to 2) $\times10^{20}$ kg of S over 4.57 Gyr. This is only 5 to 10% of its initial sulfur inventory, much lower than our calculation of 94 to 99% sulfur loss. We consider several interpretations of this difference.

It is possible that Io's initial sulfur inventory was smaller than we estimated above, for example if Io formed with lower sulfur abundance than the L/LL chondrites. However, if Io's initial sulfur abundance was instead closer to the Solar System average than to ordinary chondrites (26), it would contain more sulfur not less. Io might also have lost a substantial fraction of its initial sulfur content soon after formation (27), leaving a smaller effective reservoir for its subsequent mass loss. Another possibility is that the initial $^{34}S/^{32}S$ ratio of Io was higher than the Solar System average. The sulfur in L/LL chondrites has $\delta^{34}S_{VCDT}$ = -0.02±0.06‰ (28), consistent with the Solar System average. The most isotopically fractionated sulfur reservoirs across Earth, the Moon, Mars, and meteorites are tens of permille (Figure 4A), which is an order of magnitude less fractionated than our measurement of SO$_2$ in Io's atmosphere. The only $^{34}S$ measurements for outer Solar System material are for comets. The most precise cometary measurement was made in situ by the Rosetta spacecraft and showed that Comet 67P/Churyumov–Gerasimenko is depleted in the heavy isotope in all sulfur-bearing species (29). If some of Io's sulfur came from cometary material its initial $\delta^{34}S_{VCDT}$ would be lower than Solar System average, not higher. We therefore consider it unlikely that Io had an initial $^{34}S/^{32}S$ ratio that was much higher than the Solar System average.

Alternatively, only a fraction of the sulfur in Io might participate in the mixing and loss cycle, in particular if sulfur is concentrated in the moon's core. Io's mean density and moment of inertia indicate the presence of a core with possible compositions that range from pure Fe to an Fe-FeS eutectic mixture [~25% sulfur by mass (30)]. These constraints, combined with experimental constraints on equilibrium sulfur partitioning between metal and silicates indicate that 80-97% of Io's initial sulfur inventory is in the core (20). Our measurement of 94 to 99% sulfur loss is therefore consistent with the fraction of non-core sulfur lost if Io has been losing mass at ~0.5 to 5 times its current rate over its entire 4.57 Gyr lifetime. This implies that Io's mass loss rate could have been higher in the past than it is today.

### The $^{37}Cl/^{35}Cl$ ratio

Combining the NaCl and KCl results from both hemispheres (Table 1) gives a $^{37}Cl/^{35}Cl$ ratio of 0.403±0.028. This value is dominated by the NaCl lines, which have much higher SNR than the KCl lines. Similarly to sulfur, we convert this ratio to a $\delta^{37}Cl_{SMOC}$ value, defined as the deviation from the Earth isotopic standard, standard mean ocean chloride (SMOC), which has $^{37}Cl/^{35}Cl$ = 0.320 (31). We find $\delta^{37}Cl_{SMOC}$ = +263±88‰ for Io, which is compared to other Solar System reservoirs in Figure 4B. Chlorine participates in an outgassing and recycling process analogous to sulfur, being lost by plasma interactions at a rate of a few percent of that of sulfur (*32*). However, the $^{37}$Cl/$^{35}$Cl loss fractionation factor ($^{37}\alpha_{loss}$) is more uncertain than that of sulfur. NaCl and KCl gasses in Io's atmosphere are not replenished by sublimation, and are destroyed by photodissociation within a few hours of entering the atmosphere (*33, 34*). Their gas temperatures are high, as discussed above, and their gas dynamics appear to be dominated by volcanic plume processes. We therefore do not expect steady state gravitational stratification for these molecules, and so $^{37}\alpha_{loss}$ should be closer to 1 than $^{34}\alpha_{loss}$. Applying the same distillation reasoning as for sulfur, this leads to the conclusion that $\delta^{37}Cl_{SMOC}$ would be lower than $\delta^{34}S_{VCDT}$ for the same fractional loss. Our measured value of $\delta^{37}Cl_{SMOC}$ therefore also indicates a history of mixing and mass loss, supporting our interpretation of the $\delta^{34}S_{VCDT}$ measurement.

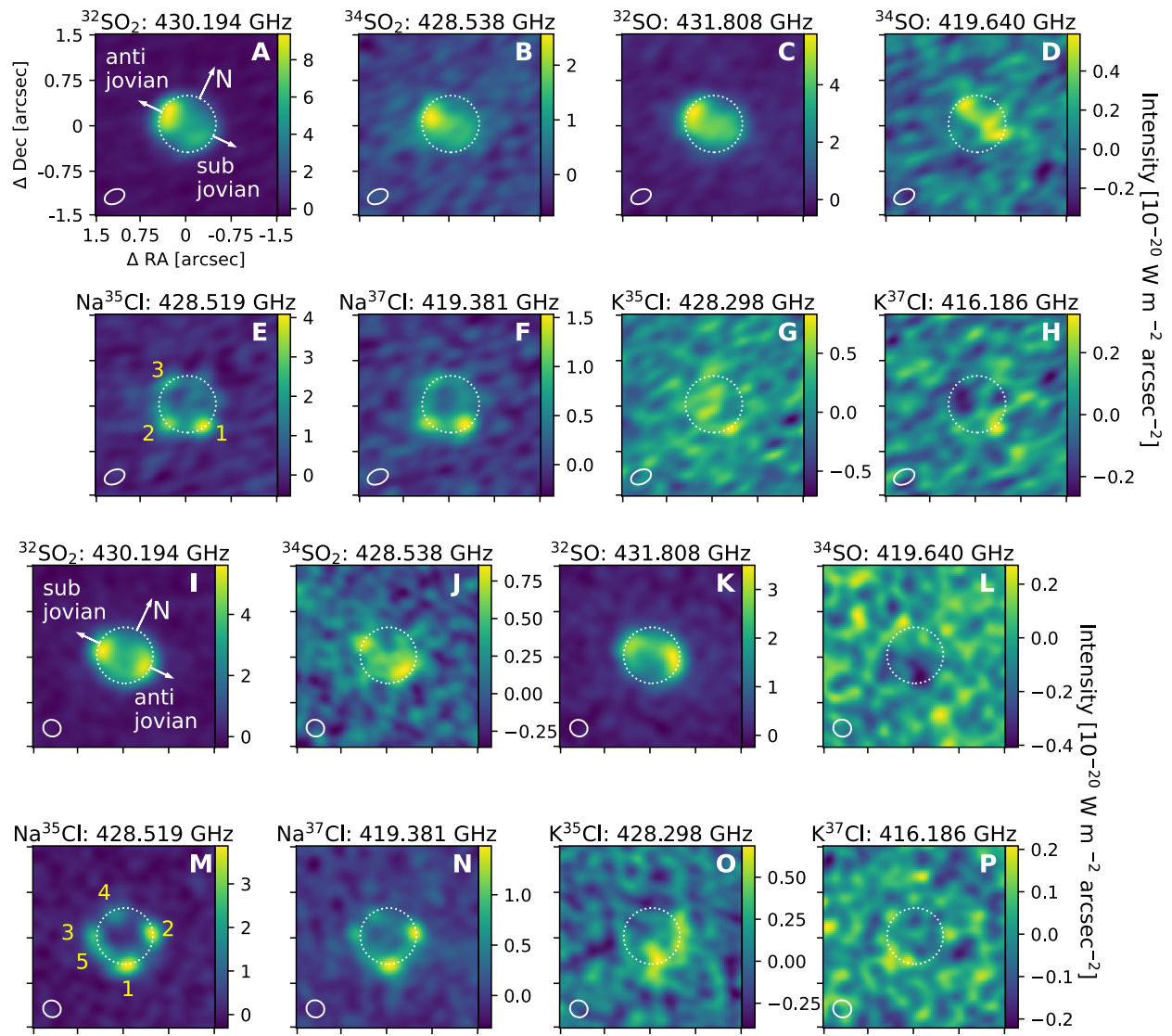

**Fig. 1. Observed distributions of molecular emission lines from Io.** Each panel shows an image extracted from the ALMA data cube for (**A-H**) the leading hemisphere and (**I-P**) the trailing hemisphere. For each species (labelled above each panel) we show the strongest detected line, on different intensity scales (color bars). Figures S1 and S2 show equivalent images for all measured lines. The dotted white circles indicate the size and location of Io, and the arrows indicate the north pole and sub- and anti-jovian directions. The white ellipses in the lower left corners indicate the size, shape and orientation of the reconstructed beam (the spatial resolution). Specific plumes are numbered in panels E and M and associated with surface features in Table S4. All x and y axes are on the scale given in Panel A in units of RA and Dec offset (Δ RA and Δ Dec, respectively). For NaCl and KCl, we interpret the discrete locations of gas emissions as plumes.

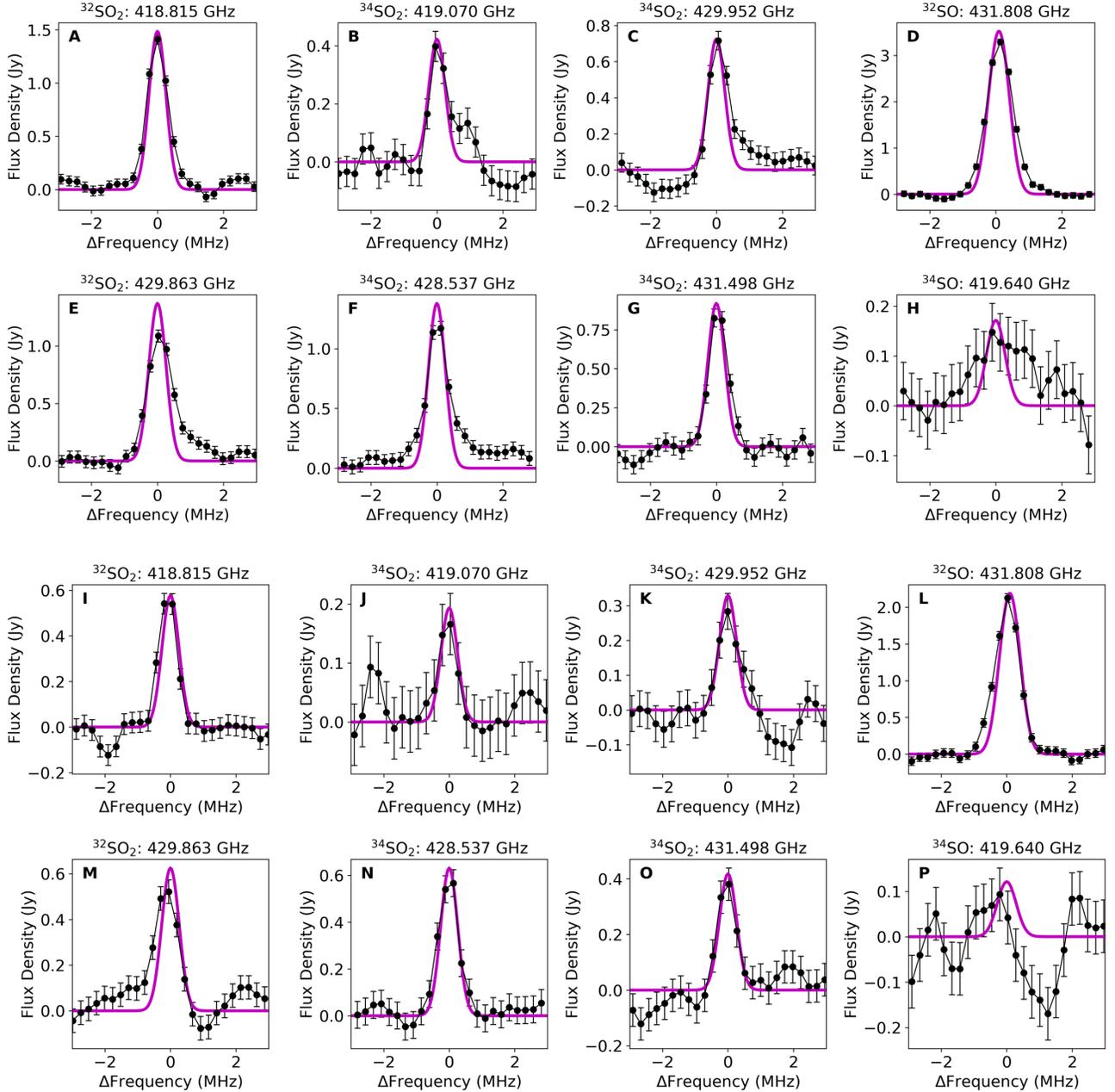

**Fig 2. Spectra of emission lines from the sulfur-bearing molecules.** Black points show the spectra in janskys (Jy) extracted from the ALMA data cube, integrated over (**A-H**) the leading hemisphere and (**I-P**) the trailing hemisphere, for each of eight emission lines (species and frequency labelled above each panel). Error bars are 1σ. Pink curves are our best-fitting atmospheric models, fitted to all the lines simultaneously, which have the parameters listed in Table 1.

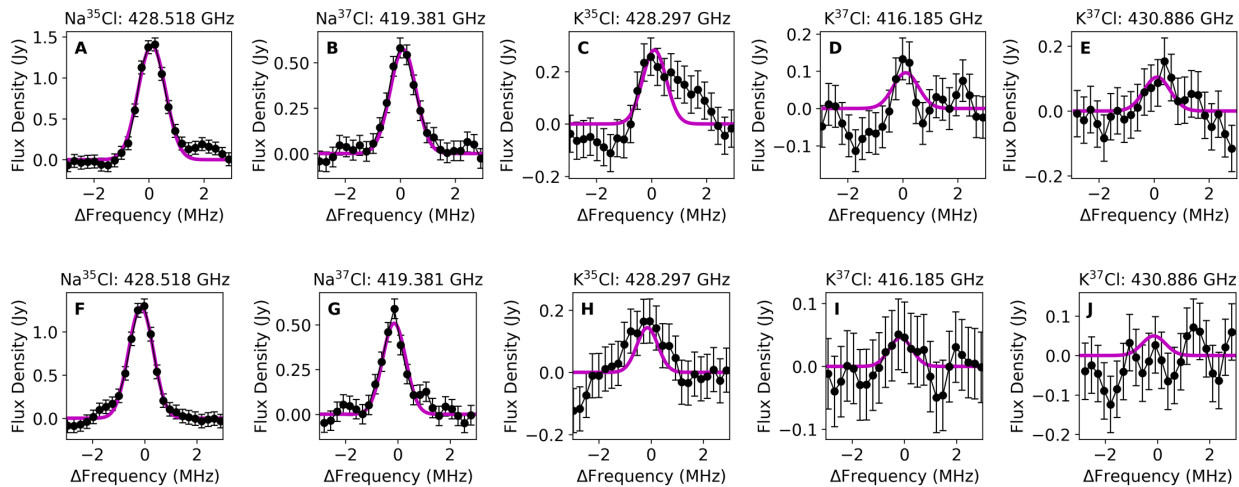

**Fig 3. Same as Fig. 2, but for the chlorine-bearing molecules.**

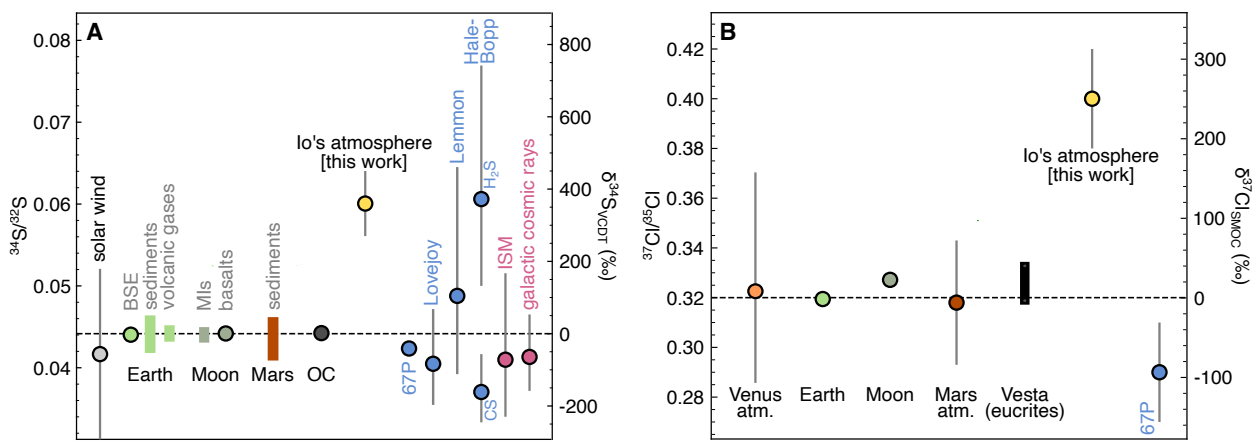

**Fig. 4. Isotopic measurements of Io's atmosphere compared to other Solar System bodies.** (**A**) Sulfur isotopes, on both the $^{34}$S/$^{32}$S and $\delta^{34}$S$_{VCDT}$ scales. (**B**) Chlorine isotopes, on the $^{37}$Cl/$^{35}$Cl and $\delta^{37}$Cl$_{SMOC}$ scales. In both panels, points indicate single measurements, while rectangular bars indicate measurements of a range of samples, and the yellow circle is our measurement for Io's atmosphere. Other points show different bodies, including bulk silicate Earth (BSE), lunar melt inclusions (MIs), ordinary chondrite (OC) meteorites, Comet 67P/Churyumov-Gerasimenko (67P), Comet C/2014 Q2 (Lovejoy), Comet C/2012 F6 (Lemmon), molecules CS and H$_2$S in Comet Hale-Bopp, the interstellar medium (ISM), the atmospheres (atm.) of Mars and Venus, and eucrites meteorites from Vesta. The x-axis is organized with distance from the Sun, not to scale. Data sources listed in Table S5 and error bars are 1σ.

**Table 1. Atmospheric model parameters.** The best-fitting model parameters determined by fitting our atmospheric models to the observed molecular emission lines. The models for sulfur- and chlorine-bearing molecules were fitted separately and have different numbers of free parameters. The observations of each hemisphere were also fitted separately. Uncertainties are 1σ (*7*). Column densities of sulfur-bearing molecules assume that the gas is uniformly distributed over Io's surface; if the assumed surface coverage were decreased, the derived column densities would increase proportionately, but other parameters would be unchanged. Similarly, the column densities of the chlorine-bearing molecules are sensitive to the emission angle that is adopted for the model calculation.

| | Sulfur-bearing molecules | | |
|---|---|---|---|
| Location | $SO_2$ column density [cm$^{-2}$] | $T_{gas}$ [K] | $^{34}SO_2/^{32}SO_2$ |
| Leading hemisphere | $(1.029 \pm 0.032) \times 10^{16}$ | 225.9±3.3 | 0.0543±0.0022 |
| Trailing hemisphere | $(3.53 \pm 0.21) \times 10^{15}$ | 240.1±7.5 | 0.0646±0.0053 |

| | Chlorine-bearing molecules | | | | | |
|---|---|---|---|---|---|---|
| | NaCl column density [cm$^{-2}$] | KCl column density [cm$^{-2}$] | Fractional coverage | $T_{gas}$ [K] | $^{37}Cl/^{35}Cl$ | Velocity [m s$^{-1}$] |
| Leading Hemisphere | $(5.1 \pm 2.0) \times 10^{13}$ | $(9.9 \pm 3.9) \times 10^{12}$ | 0.133±0.048 | 774±66 | 0.415±0.026 | 75±12 |
| Trailing Hemisphere | $(3.3 \pm 1.8) \times 10^{13}$ | $(3.5 \pm 2.0) \times 10^{12}$ | 0.158±0.071 | 682±56 | 0.391±0.029 | -99.0±5.0 |

**Acknowledgments:** We thank Arielle Moullet for insight into past observations of Io, and Alexander Thelen for help with CASA imaging. We acknowledge the support of Ryan Loomis, Tony Remijan, and the North America ALMA Science Center (NAASC) in obtaining these data and processing them into calibrated images. This project concept was developed in part at the W.M. Keck Institute for Space Studies. ALMA is a partnership of ESO (representing its member states), NSF (USA) and NINS (Japan), together with NRC (Canada), MOST and ASIAA (Taiwan), and KASI (Republic of Korea), in cooperation with the Republic of Chile. The Joint ALMA Observatory is operated by ESO, AUI/NRAO and NAOJ. The National Radio Astronomy Observatory is a facility of the National Science Foundation operated under cooperative agreement by Associated Universities, Inc. The Jet Propulsion Laboratory (JPL) is operated by the California Institute of Technology under contract with the National Aeronautics and Space Administration (80NM0018D0004).

**Funding:**

KdK acknowledges funding from National Science Foundation grant 2238344 through the Faculty Early Career Development Program.

KdK, JE, EH, and AEH acknowledge funding from the Caltech Center for Comparative Planetary Evolution.

KM acknowledges support from NASA ROSES Rosetta Data Analysis Program grant 80NSSC19K1306.

AEH acknowledges support from the JPL Researchers on Campus Program and from internal JPL funding.

KdK acknowledges support from the NAASC through their funding of a PI face-to-face data reduction visit.

**Author contributions:** KdK led the conceptualization, methodology development, analysis, interpretation, and writing of the manuscript. FN, JE, and KM contributed to project conceptualization and to development of the Rayleigh distillation model. AEH and EH contributed to development of the Rayleigh distillation model. SLC developed the radiative transfer model with contributions from KdK. All authors contributed to interpretation and to review and editing of the paper.

**Competing interests:** The authors declare that they have no competing interests.

**Data and materials availability:** This paper makes use of the following ALMA data: ADS/JAO.ALMA#2021.1.00849.S, which archived at https://almascience.nrao.edu/aq/?projectCode=2021.1.00849.S . The radiative-transfer modeling software is archived at Zenodo (*8*) doi:10.5281/zenodo.10794511. Measured atmospheric parameters are listed in Tables 1 and S3.

**Supplementary Materials**
Materials and Methods
Supplementary Text
Figs. S1 to S6
Tables S1 to S5
References *(35-79)*

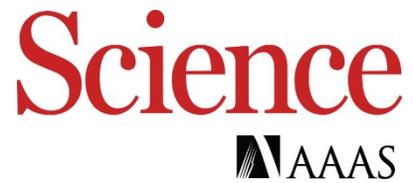

Supplementary Materials for

**Isotopic evidence of long-lived volcanism on Io**

Katherine de Kleer[1]*, Ery C. Hughes[1,2], Francis Nimmo[3], John Eiler[1], Amy E. Hofmann[4], Statia Luszcz-Cook[5,6,7], Kathy Mandt[8].

*Corresponding author, email: dekleer@caltech.edu

**The PDF file includes:**

    Materials and Methods
    Supplementary Text
    Figs. S1 to S6
    Tables S1 to S5

**Materials and Methods**

Observations

Observations were made with ALMA of Io's leading and trailing hemispheres on UT 2022 May 24 and 2022 May 18 (respectively) through program 2021.1.00849.S. QSO B0420-0127 and QSO B2251+155 were observed for flux and bandpass calibration. The time on source was 1h40m for the leading hemisphere observation, and 50 minutes for the trailing hemisphere observation, because better weather conditions allowed the desired SNR to be obtained in a shorter time. Parameters of the observations and Io's geometry at the time are given in Table S1. For the leading hemisphere observation, Io rotated 28 degrees during the observation, which smears the data by 0″.23 in the rotation direction; this is more than half the spatial resolution, but does not affect our interpretation because all presented spectra are averaged over regions larger than the spatial resolution.

The observations were conducted while ALMA was in its C-4 antenna configuration, which provides a spatial resolution of about a quarter to a third of Io's diameter (~1000 km) at the frequencies of observation. The maximum baselines were 780 and 740 m for the leading and trailing hemisphere observations respectively.

The observations used ALMA's Band 8 receivers, operating between 416 and 432 GHz. Within this frequency coverage we selected 13 spectral windows, each with a frequency resolution of 244 kHz (corresponding to a velocity resolution of 170 m s$^{-1}$) and a bandwidth of either 235 or 118 MHz, for a total recorded bandwidth of 1.5 GHz. Band 8 covers rotational lines of all 4 targeted species ($SO_2$, SO, NaCl, KCl) and their isotopologues in a single spectral set-up (Table S2). ALMA's frequency tunings were set to track the changing line-of-sight velocity of Io during the observations, so there is no spectral smearing provided the tracking uses an ephemeris sampled at sufficient precision. For these observations the ephemeris (*35*) was sampled at 10 minute intervals; the resulting spectral smearing is < 2 kHz, which is a small fraction of the spectral resolution. The full width at half maxima for the observed lines are in the range of 0.6 to 1.0 MHz for the sulfur-bearing species and 0.95-1.15 MHz for the chlorine-bearing species. A tuning error resulted in some of the spectral windows being tuned incorrectly for the leading hemisphere observation such that two $SO_2$ lines (430.229 and 430.232 GHz) were not covered by that observation. These lines are not used in our analysis.

Data reduction

The data were processed through the standard ALMA pipeline (*36*) to produce a calibrated measurement set (MS) containing the interferometric visibilities, which are the amplitude and phase of the cross-correlated signal between each pair of antennas. The line-free spectral channels across all spectral windows were split out to produce a continuum MS. This continuum MS was self-calibrated (*37*) and imaged with the CASA software (*38, 39*) using an iterative procedure. First, a continuum limb-darkened disk the size of Io at the time of observation was produced and converted to visibilities; a limb darkening parameter of 0.2 was used, but the final image is only weakly sensitive to the choice of limb darkening parameter. The data were then phase calibrated using the limb-darkened disk model visibilities, and imaged using the same model as a starting point for the CLEAN deconvolution algorithm (*40*). The self-calibration and CLEAN deconvolution process was then repeated, using the output of the previous self-calibration and imaging round as the starting point for the subsequent iteration. In each round, the self-calibration was performed using an increasingly shorter solution interval, and the CLEAN algorithm was employed using an increasingly deep threshold. We used three iterations of self-calibration and deconvolution; further iterations did not improve the SNR of our images.

We applied the phase calibration derived from the continuum data (described above) to the continuum-subtracted spectral line data, then imaged each channel in the spectral line data channel separately. This produced a spectral data cube for each spectral window with ~0″.28 resolution and a spectral sampling of 244 kHz, and a frequency-averaged continuum image at ~0″.28 resolution. Figures S1 and S2 show the continuum-subtracted image integrated over each spectral line listed in Table S2. We find that each spectral line from a single species, including both isotopologues, has the same spatial distribution, which indicates that the measured spatial distributions of the species are not biased by artifacts.

The 1D spectrum in each spectral window was extracted using an aperture that includes all pixels that are 5% of the peak continuum level or higher in the continuum image. This produces an aperture whose diameter is 1.3× the diameter of Io, i.e., extending one resolution element beyond the edge of Io.

The flux density uncertainties were estimated in two ways. First, a 1D spectrum was extracted in a region absent of sources, in exactly the same way as for the source and using the same aperture size; the standard deviation of the spectrum across each spectral window was used as an estimate of the noise in that spectral window. Second, the standard deviation of the spectrum of Io in spectral regions without apparent spectral lines was calculated, again providing a noise estimate per spectral window. We adopt the more conservative uncertainties derived from the latter method, because it incorporates the thermal noise from Io's continuum. The first method gave uncertainties that were a factor of 1 to 3 lower, which we regard as under-estimates. The degree of bandpass calibration noise varies between spectral windows, and is non-negligible in some windows. The bandpass calibrator spectrum was smoothed using a frequency width of 7.8 MHz. This smoothing reduces the noise introduced by the bandpass calibrator, and although it can introduce spectral artifacts, such artifacts would be similar in width to the smoothing window and hence much broader than Io's emission lines. The uncertainties on the datapoints incorporate this noise, which is particularly high in the spectral windows containing the lines at 419.640, 428.298, 429.863, 429.952, and 420.887 GHz. The 1σ noise is shown in Figures 2 and 3 and incorporated into the maximum-likelihood calculations (see below).

Overview of modeling and retrievals

In order to determine the isotope ratios for sulfur and chlorine, we found best-fitting model parameters by fitting a forward model to the observations as follows. Model spectra were generated using a radiative-transfer model for the atmosphere of Io (*8, 9*), which we updated to add additional species. Our model includes opacity from $SO_2$, SO, NaCl, and KCl, including the $^{32}S$ and $^{34}S$ isotopes of sulfur and the $^{35}Cl$ and $^{37}Cl$ isotopes of chlorine. Models were fitted to the data using a Nelder-Mead minimization algorithm as implemented in the `optimize` package in the SCIPY software (*41*). Once the best-fitting solution was determined, we used the EMCEE Markov chain Monte Carlo (MCMC) Ensemble sampler (*42*) to explore the model parameter space and determine the uncertainties on the best-fit parameters. Additional details on each of these steps are given below.

The parameter values that correspond to the maximum likelihood values output by the MCMC simulations match the best-fitting values found by the minimization, with differences well below 1σ. We report the resulting parameter uncertainties as the 1σ range measured from the posterior probability distributions output by the MCMC simulations. Table 1 reports the best-fitting parameters and MCMC-derived uncertainties for all free parameters in the models.

Emission line selection and treatment

For the radiative-transfer modeling, the line frequencies and strengths were adopted from the Cologne Database for Molecular Spectroscopy [CDMS (*43, 44*)]. We also tested line lists from the JPL Molecular Spectroscopy repository (*45*), but found they did not match the observed line positions in our datasets for any species except $^{32}SO_2$ and $^{34}SO_2$. The CDMS frequencies for the lines of the sulfur-bearing molecules agree with our observations after accounting for the line-of-sight velocity of Io. For NaCl, KCl, and their isotopologues, we observed an additional velocity shift that we ascribe to bulk motion of the gas and included as a free parameter in our model fitting. The observed ~100 m s$^{-1}$ velocity shift is the same across all chlorine-bearing species within each dataset; it is smaller than the velocities of Io's largest class of plumes [500-1000 m s$^{-1}$ (*46*)]. A velocity shift parameter is not included in the $SO_2$ model because the line positions match the expected frequencies from CDMS. The velocity difference between the chlorine- and sulfur-bearing species could arise because $SO_2$ and SO are more uniformly distributed, such that gas velocity components produce line broadening rather than a frequency shift. The broadening of the $SO_2$ and SO lines due to Io's rotational motion is included in the model (see below). The velocity shifts for the chlorine-bearing gasses are unlikely to be due to line list errors, because the shift is the same across NaCl and KCl lines (per dataset), whereas line list errors introduce offsets that differ between lines but are the same between datasets (per line). The frequency errors that would be introduced by using the JPL line lists are much larger than the observed frequency shifts due to gas velocity.

The strengths of the emission lines are sensitive to temperature, and the temperature profile in Io's atmosphere is poorly known. The isotope ratio also varies with altitude due to gravitational stratification. To determine the isotope ratio, we therefore used only the two lines of $^{32}SO_2$ (418.815 and 429.863 GHz) that are sensitive to the same low atmospheric altitudes as the $^{34}SO_2$ lines. The other $^{32}SO_2$ lines covered by our data, as well as those used for the previous isotope ratio measurement (*15*), have higher line opacities such that emission arises predominantly from higher altitudes than the $^{34}SO_2$ lines are primarily sensitive to. This is particularly true near the limb where the path length through the atmosphere is longest. This is illustrated in Figure S3, which shows the contribution functions calculated from the model opacities at line center for the $^{32}SO_2$ and $^{34}SO_2$ lines targeted in our observations, compared to those used for the previous isotope ratio measurement. Particularly near the limb, all the $^{32}SO_2$ lines used in the past work, and most of the $^{32}SO_2$ lines covered by our data, are sensitive to different altitudes than the $^{34}SO_2$ lines.

The $SO_2$ gas temperature is tightly constrained by our observations because the $SO_2$ model fitting includes six emission lines from high to low excitation (Table S2). Temperature affects both the line widths and the relative strengths of the lines. If each line were fitted independently, the temperature would be degenerate with column density and with the velocity distribution of the gas, leading to much larger uncertainties. However, because we fit multiple lines simultaneously, the best-fitting temperature is constrained by the relative line strengths. The resulting best-fitting temperature, in combination with Io's rotation, then determines the model line widths. The imperfect match between the model line widths and some of the observed spectra (Fig. 2) could arise from velocity components (e.g. from winds or plumes) that are not included in our model.

Model atmosphere geometry and line opacity

To determine a disk-integrated model spectrum for comparison with the data, we modeled the emission from Io's atmosphere as a function of latitude and longitude, accounting for the dependence of atmospheric path length on emission angle and the Doppler shift corresponding to Io's solid-body rotation at each latitude and longitude. The model was output with a range of spatial resolutions, from which we selected the coarsest model resolution that did not result in a disk-integrated spectrum that differed substantially from that produced by higher resolution models. Based on this criterion, we selected models generated with a spatial resolution of 0″.06 (about 6% of Io's diameter), which were then spatially integrated to produce a disk-integrated model spectrum. Changing the spatial resolution of the model causes minor changes in the derived column densities, but does not affect the derived isotope ratios.

For the $^{34}S/^{32}S$ model fitting, we assumed that Io's atmosphere is homogeneous in latitude and longitude. Figure 1 indicates that the observed fractional coverage (fraction of the surface area of Io above which there is $SO_2$ gas) is closer to ~50%. If the lines are optically thin, there is a linear trade-off between column density and fractional coverage: if column density is increased and fractional coverage decreased proportionally, the model line strength remains the same. This is the case provided the fractional coverage is above ~15%. Our assumption of uniform coverage therefore does not impact the derived isotope ratio, but it does affect the derived column densities.

For NaCl and KCl, the fractional coverage is unclear from the images. Figure S3 shows contribution functions that have the same disk-integrated column for the cases of 20% and 5% fractional coverage, demonstrating that if these species exhibit a lower fractional coverage and higher column density, the observed emission is coming from higher altitudes than if the species are more uniformly distributed across Io. If the fractional coverage is below ~10% there is enough opacity in the $Na^{35}Cl$ line that it no longer traces emission from the same altitudes as the $Na^{37}Cl$ line. This necessitates our inclusion of fractional coverage as a free parameter in the chlorine model fitting; if a broad range of fractional coverages is allowed by the data, the effect is to increase the derived uncertainties on all parameters that are correlated with fractional coverage. In our analysis, including this free parameter primarily increases the uncertainties on the gas column densities, because the best-fitting models are in a region of the parameter space where opacity is low and the derived isotope ratio is not strongly impacted.

As an additional check of whether opacity effects may bias the derived isotope ratios, we extracted spectra from localized regions on Io's disk with both low and high path lengths (disk center and low-latitude limbs, respectively) then applied the sulfur model fitting. For the chlorine model fitting, we performed the same check using spectra from both fainter and brighter emission regions. These tests used circular apertures with 0″.3 diameters, shown in Fig. S4, to extract the spectra. The best-fitting parameters for each region are given in Table S3. This test assumes a coverage fraction of 1.0 within the 0″.3 aperture. Some of the NaCl and KCl emission occurs very close to the limb. We find that the derived column densities are sensitive to the exact emission angle used in the models, so should be interpreted with caution, but the derived isotope ratios are not sensitive to this parameter. The results of this test show that the $^{34}S/^{32}S$ and $^{37}Cl/^{35}Cl$ ratios derived from all spectra extracted within a given hemisphere differ by <1.5σ from the values derived from the disk-integrated spectra. The lack of systematic differences in the isotope ratios derived for low and high emission regions indicates that local opacities do not bias our derived isotope ratios. For the best-fitting atmospheric parameters and model resolution adopted above, the optical depths of all lines are <0.1 over most of the surface, and always <0.5 for the sulfur-bearing species and <0.7 for the chlorine-bearing species (with the highest values being for the strongest lines at the limbs).

The lines of $Na^{35}Cl$ and $Na^{37}Cl$ are detected at a much higher SNR than $K^{35}Cl$ and $K^{37}Cl$, due to the higher abundance of NaCl relative to KCl. Therefore the chlorine model fitting is dominated by NaCl; using the $Na^{35}Cl$ and $Na^{37}Cl$ lines alone gives the same $^{37}Cl/^{35}Cl$ ratio, within the 1σ uncertainties, as using both NaCl and KCl. The Na/K ratios for both hemispheres, including the disk-integrated and local analyses, are all in the range 3 to 10, consistent with previous studies (*14*).

<u>Examples of derived uncertainties</u>

Figure S5 shows a random selection of model spectra drawn from the posterior probability distribution, compared to the observed sulfur data for the trailing hemisphere, to illustrate the variations in the spectra produced by varying the parameters within their uncertainties. Figure S6 shows models in which the $SO_2$ column density is fixed at its best-fitting value but the $^{34}S/^{32}S$ ratio is varied from 0.040 (below the Solar System average) to 0.080 (above our best-fitting value).

**Supplementary Text**

<u>Patera co-located with chlorine-bearing gasses</u>

Our models indicate that NaCl and KCl have high gas temperatures, and Fig. 1 shows they are localized to discrete locations. As discussed in the main text, we interpret them as only present in volcanic plumes. For each of the two dates of observation, we determined the position of each source in the NaCl images and converted it to a latitude and longitude on Io using the geometry at the time of observation (Table S1). The latitudes and longitudes of the sources marked in Fig. 1 are given in Table S4. The uncertainties on the latitudes and longitudes were determined by calculating the latitude and longitude of every pixel within a 10×10 pixel box (roughly one resolution element on each side) surrounding the determined source center, then taking the standard deviation within the box. We investigated whether these correspond to known surface features (*47*). Table S4 lists the most likely patera as well as all paterae that fall within the 1σ uncertainties.

Volcanic activity at Kurdalagon Patera (consistent with location 1 in Figure 1), is thought to have been at least partially responsible for the massive brightening of Jupiter's sodium nebula in early 2015 (*48*); our identification of NaCl gas at the location of the patera is consistent with this connection because it suggests that the style of volcanism taking place at Kurdalagon Patera produces Na-bearing gas. However, infrared images taken simultaneously with our ALMA observations on 2022 May 24 (*49*) do not show thermal emission at the latitudes and longitudes where we observe NaCl and KCl gasses. If the regions of high NaCl and KCl gas density are indeed volcanic plumes, they originate from volcanic centers that are not actively extruding large volumes of lava.

Rayleigh distillation model

We use a Rayleigh distillation model to relate the present-day ratio of two isotopes to the fraction of material that has been lost from the system over time. This relationship is quantified through the Rayleigh equation, which is $^{34}R = {}^{34}R_0 f^{{}^{34}\alpha_{loss}-1}$ for the $^{34}$S and $^{32}$S sulfur isotopes. We use this to determine the value of *f*, the fraction of Io's original sulfur inventory remaining at present day, based on our measured $^{34}R$.

Rayleigh fractionation entails the progressive and irreversible removal of material from a system (referred to as a reservoir). We assume an initial $^{34}$S/$^{32}$S ($^{34}R_0$) for our system and a value for the fractionation factor ($^{34}\alpha_{loss}$), which describes the instantaneous isotopic partitioning between the modelled system and each packet of material removed at each time step. The Rayleigh distillation framework assumes that i) the material is being removed continuously, and ii) the residue is well mixed.

For the case of sulfur on Io, we propose that the well-mixed system consists of all of Io's sulfur that is not in the moon's core. It is also possible that a smaller shallow system, consisting just of Io's atmosphere and crust (and perhaps some portion of the upper mantle), is mixed more rapidly, due to shallow re-melting of surface frosts recycled into the crust. In such a scenario, this smaller well-mixed sulfur reservoir would become isotopically fractionated more rapidly than the full mantle plus crust system. However, to maintain the crust-atmosphere reservoir, mantle material would need to be continuously injected to balance Io's mass loss rate of 1000 to 3000 kg s$^{-1}$. This injection of essentially unfractionated mantle material would buffer the near-surface system such that the atmospheric $^{34}$S/$^{32}$S could not become highly fractionated: the atmospheric $^{34}$S/$^{32}$S would take values between those of steady-state and Rayleigh fractionation scenarios (*20*). Fractionation in the shallower, smaller system could only produce a highly fractionated atmospheric $^{34}$S/$^{32}$S if there is no addition of unfractionated mantle. We consider that scenario highly unlikely, given the observed mantle-derived volcanism and the amount of sulfur input into the crust that is required to balance Io's mass loss.

By assuming the well-mixed reservoir consists of all Io's sulfur that is not in the core, our model calculation also requires that there is no sulfur exchange between the core and mantle. The sulfur isotopic fractionation factor between metal and silicates is close to 1 (*50*), so we expect that the formation of Io's core left it with the moon's initial sulfur isotope composition. If Io's core supplies sulfur to the mantle, this provides a source at Io's initial $^{34}$S/$^{32}$S ratio and therefore lowers the average $^{34}S/^{32}S$ ratio of the sulfur that is available for loss. This would therefore require even greater sulfur loss to explain our measurement.

The derived fraction of sulfur lost from Io depends on our adopted initial isotope ratio $^{34}R_0$ and $^{34}\alpha_{loss}$. As discussed in the main text, potential deviations in the $^{34}S/^{32}S$ ratio of Io-forming material from the Solar System average are expected to be four orders of magnitude smaller (in $\delta^{34}S_{VCDT}$) than our observed fractionation. Adopting the Solar System average for $^{34}R_0$ is therefore not a large source of uncertainty. The value we adopt for $^{34}\alpha_{loss}$ assumes that all loss takes place at or above an exobase located at 600 km altitude. The altitude of the exobase is uncertain and could be as low as 100 to 200 km (*51, 52*). During Io's night time, the exobase might be at the surface itself. If we adopted an exobase at a lower altitude than 600 km, it would result in an $^{34}\alpha_{loss}$ value closer to 1. This would put our loss fraction derived from the Rayleigh equation at the upper end of our reported range but does not qualitatively change our conclusions.

In our implementation of the Rayleigh model, we assume a constant fractionation factor, $^{34}\alpha_{loss}$. On Io, however, $^{34}\alpha_{loss}$ probably changes on diurnal, seasonal, and stochastic timescales, as Io's exobase altitude changes in response to changing atmospheric densities. We made the simplifying assumption of a fixed altitude because we expect the exobase is typically at or below our adopted value, such that any deviation from our assumption would result in the same conclusion, or even greater sulfur loss.

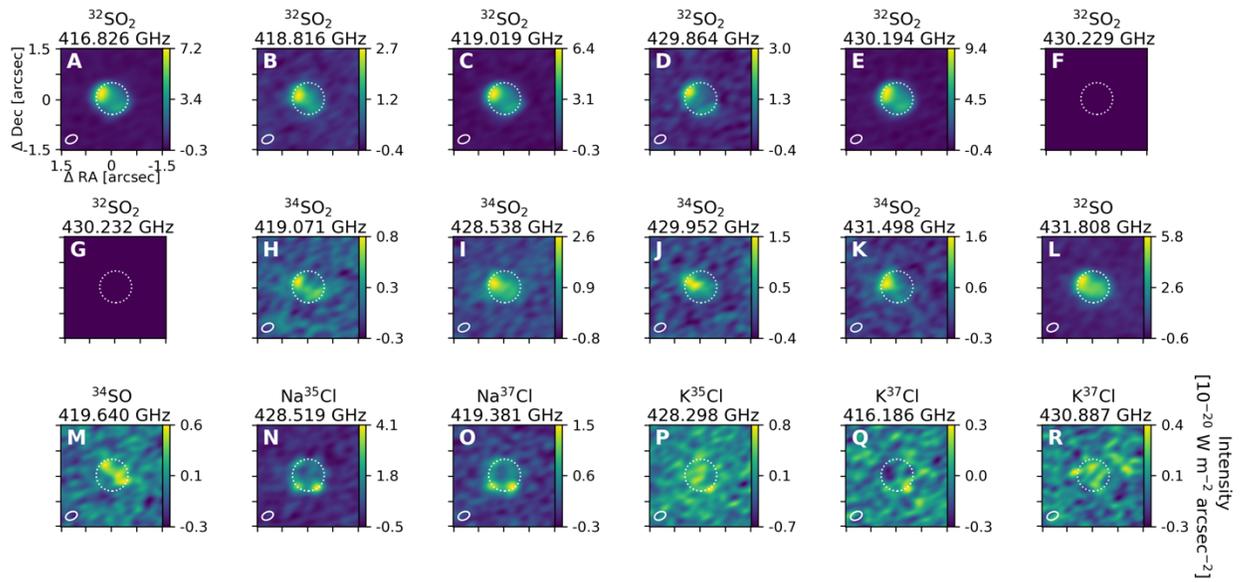

Fig. S1. **Distributions of all observed molecular emissions from Io's leading hemisphere.** Same as Figure 1, but for all the lines we observed (listed in Table S2) on the leading hemisphere for (**A-G**) $^{32}SO_2$, (**H-K**) $^{34}SO_2$, (**L**) $^{32}SO$, (**M**) $^{34}SO$, (**N**) $Na^{35}Cl$, (**O**) $Na^{37}Cl$, (**P**) $K^{35}Cl$, and (**Q-R**) $K^{37}Cl$. A tuning error led to no recorded data for the $SO_2$ lines at (**F**) 430.229 and (**G**) 430.232 GHz. Colorbars are in the intensity units given in Panel R.

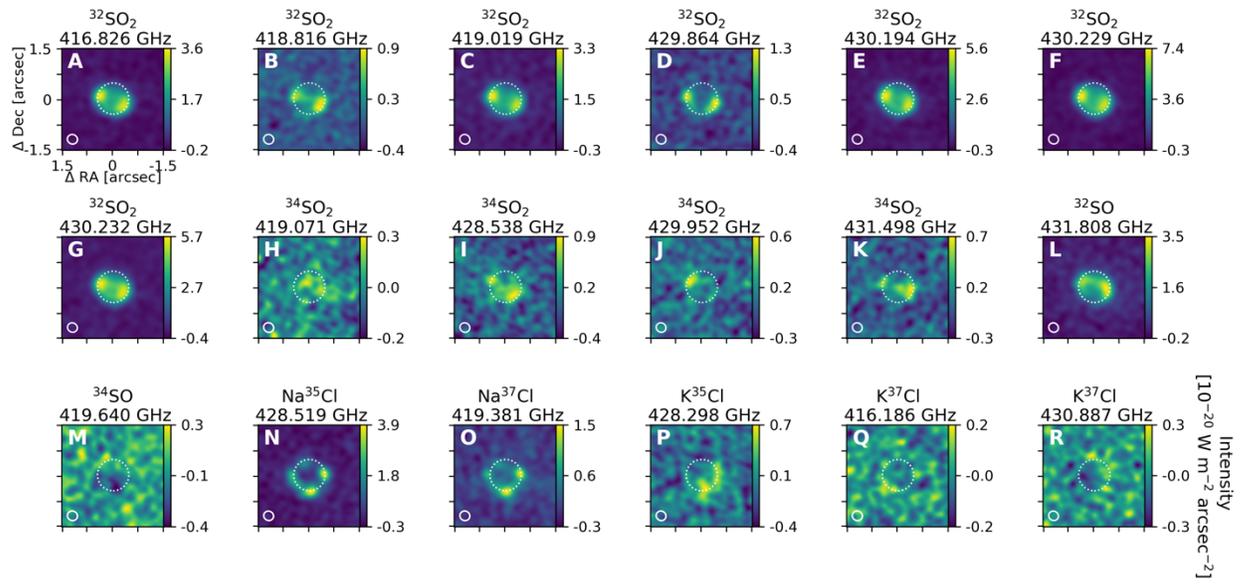

Fig. S2. **Distributions of all observed molecular emissions from Io's trailing hemisphere.** Same as Fig. S1, but for the trailing hemisphere.

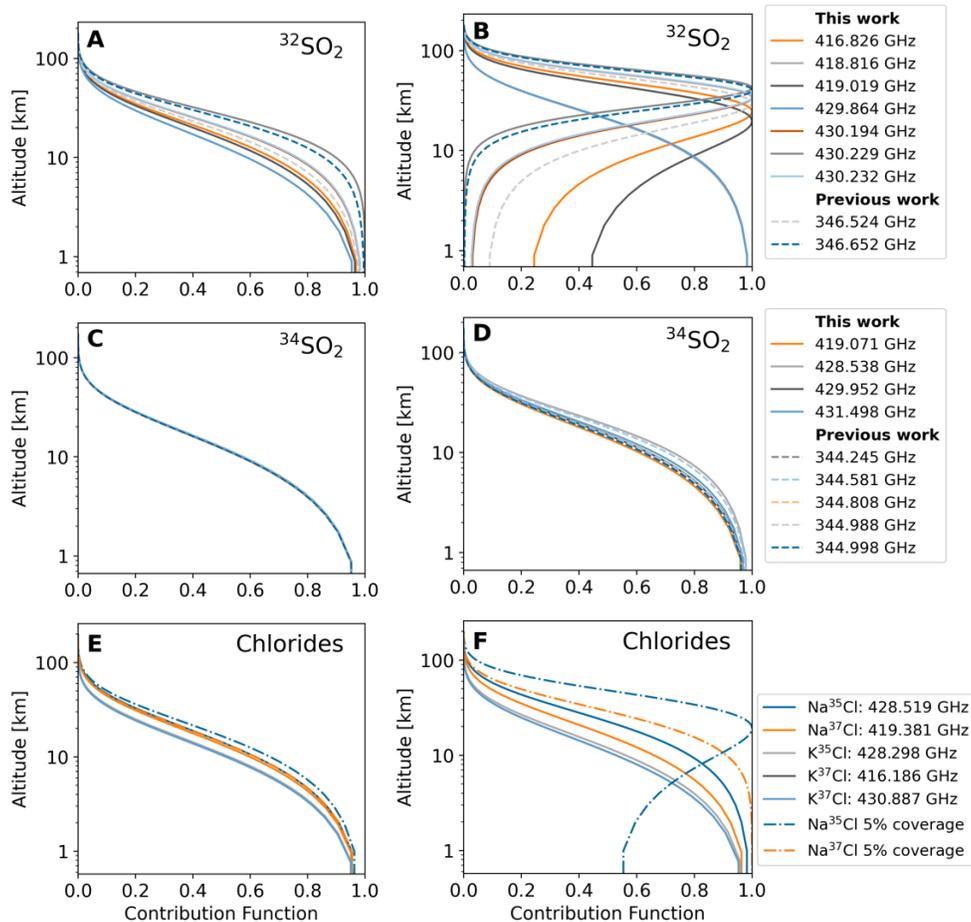

Fig. S3. **Contribution functions for emission lines from this and previous work.** Contribution functions for all observed lines of the species used to derive the sulfur and chlorine isotope ratios. The lines used in previous work (*15*) are shown for comparison. (**A,C,E**) Calculated values for the center of the disk and (**B,D,F**) the limb, for the species indicated on the panels. The $SO_2$ contribution functions assume an $SO_2$ column density of $1\times10^{16}$ cm$^{-2}$ and gas temperature of 240 K, corresponding to the leading hemisphere best fitting values in Table 1. For $^{32}SO_2$, only the lines at 418.816 and 429.864 GHz are used to derive the isotope ratio because they are sensitive to the same atmospheric altitudes as the $^{34}SO_2$ lines, especially near the limb where much of the emission appears. The NaCl and KCl contribution functions assume a temperature of 800 K and 20% fractional coverage unless otherwise indicated; some contribution functions assume an alternative 5% coverage (for an equivalent disk-integrated column) to show how much the fractional coverage can impact the relative altitudes the isotopologues are sensitive to.

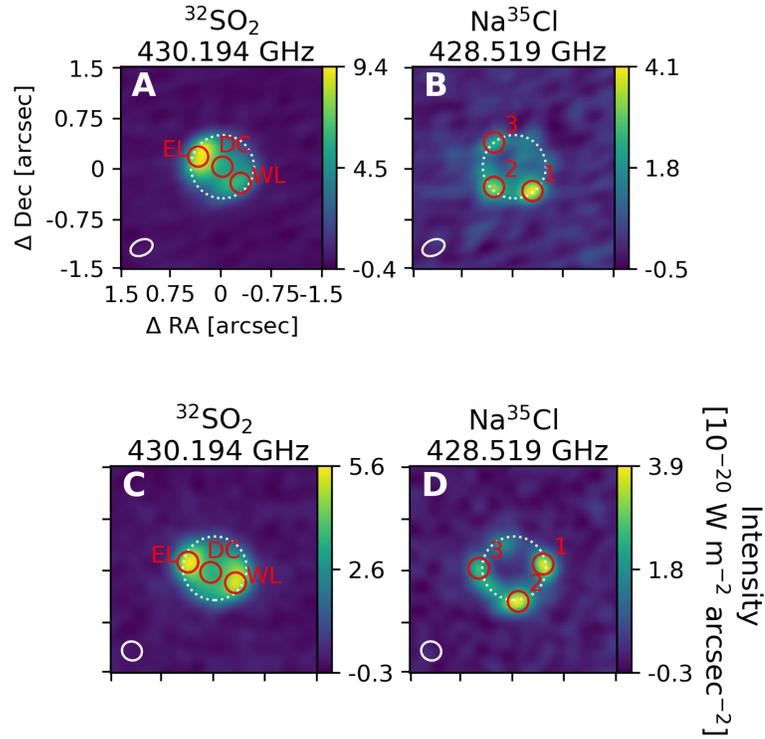

Fig S4. **Locations used for local model fitting.** Images of (**A-B**) the leading and (**C-D**) trailing hemisphere in example emission lines of (**A,C**) $^{32}SO_2$ and (**B,D**) $Na^{35}Cl$ (frequencies labeled above each panel). Red circles indicate the apertures used for our tests (see text). For $^{32}SO_2$, the identified regions are the east limb (EL), west limb (WL), and disk center (DC). For $Na^{35}Cl$, the identified regions are the three brightest emission locations in each observation. Other symbols are the same as in Figure 1.

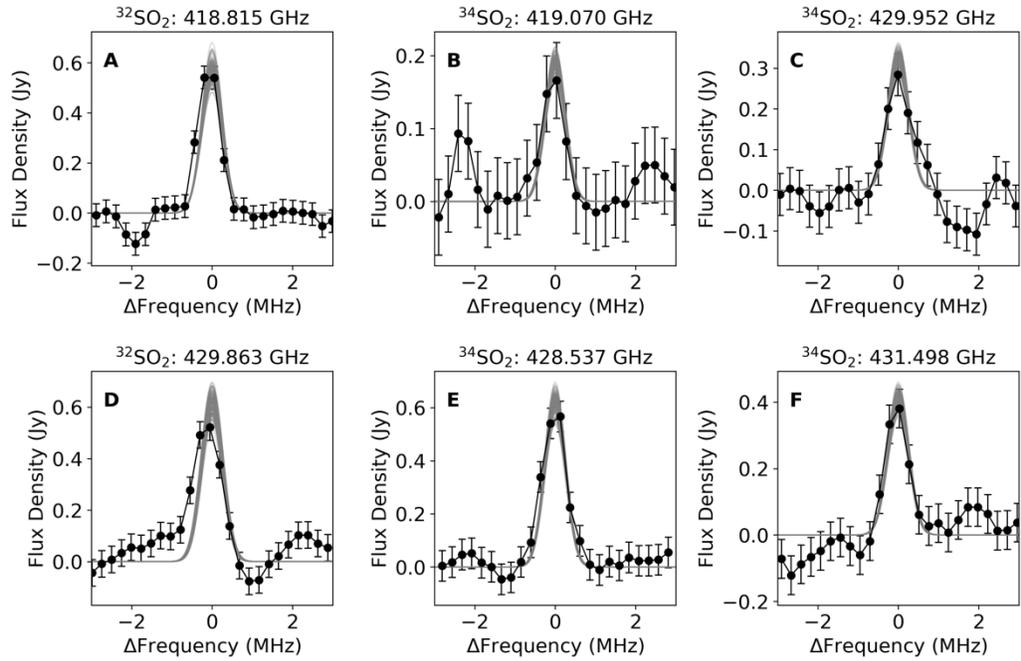

Fig S5. **Visualization of parameter uncertainties.** Same as Fig. 2I-K and M-O but with multiple models (gray curves) corresponding to 150 parameter combinations randomly selected from the joint posterior probability distribution determined by the MCMC simulation.

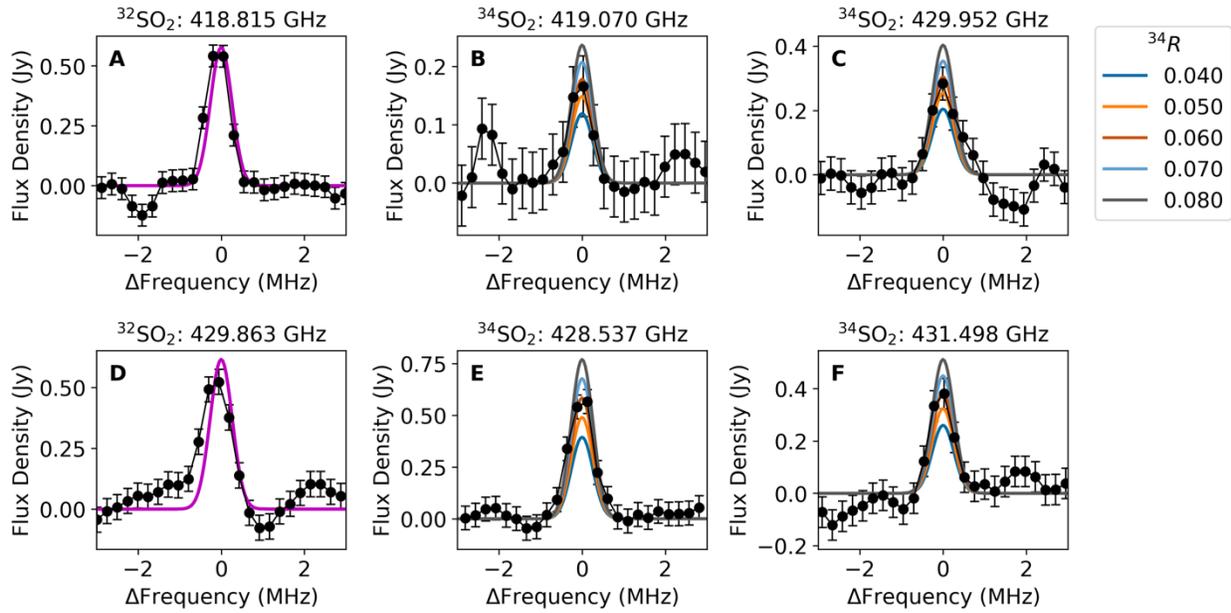

Fig S6. **Visualization of different isotope ratios.** Same as Fig. S5, but with models corresponding to different isotope ratios (see legend). The models shown all adopt the best-fitting $^{32}SO_2$ column density for this observation. The $^{34}R$ values shown for $^{34}SO_2$ were chosen to be ±5σ from the best fitting value.

**Table S1. Observing parameters for the two observations presented in this paper.**

| Hemisphere | Leading | Trailing |
|---|---|---|
| Date/Time [UT] | 2022-05-24 12:48 to 16:07 | 2022-05-18 10:28 to 12:02 |
| Time on Source | 1h40m | 50m |
| Precipitable water vapor [mm] | 0.7 to 0.8 | 0.5 |
| Angular resolution | 0″.23×0″.35 | 0″.27×0″.29 |
| Angular Diameter | 0″.937 | 0″.924 |
| Sub-obs longitude [°W] | 73 to 101 | 27 to 287 |
| Sub-obs latitude [°N] | 2.1 | 2.1 |
| North pole angle | 335° | 335° |

**Table S2. Molecular data for all emission lines detected in our observations.** Quantum numbers (QN) are given for the total rotational quantum number (J), the total rotational angular momentum (N), and the projections of N onto the A and C inertial axes ($K_a$ and $K_c$ respectively), for the upper and lower state. The lower state energy ($E_L$) is given in cm$^{-1}$. Data from CDMS (*43, 44, 53-61*).

| Species | Frequency [MHz] | $E_L$ [cm$^{-1}$] | Line Strength at 300 K [cm$^{-1}$/(molecule × cm$^{-2}$)] | QN | Detected Leading | Detected Trailing |
|---|---|---|---|---|---|---|
| $^{32}SO_2$ | 416825.5576±0.0019 | 289.053 | 6.87986×10$^{-22}$ | J=28 $K_a$=4←5 $K_c$=24←23 | Y | Y |
| $^{32}SO_2$ | 418815.8002±0.0020 | 192.736 | 1.03748×10$^{-22}$ | J=18←17 $K_a$=7←8 $K_c$=11←10 | Y | Y |
| $^{32}SO_2$ | 419019.0378±0.0019 | 331.436 | 5.58444×10$^{-22}$ | J=31 $K_a$=3←4 $K_c$=29←28 | Y | Y |
| $^{32}SO_2$ | 429863.8467±0.0019 | 659.183 | 1.91637×10$^{-22}$ | J=44 $K_a$=3←4 $K_c$=41←40 | Y | Y |
| $^{32}SO_2$ | 430193.7070±0.0015 | 180.632 | 1.05604×10$^{-21}$ | J=23←24 $K_a$=2←1 $K_c$=22←23 | Y | Y |
| $^{32}SO_2$ | 430228.6487±0.0016 | 168.493 | 1.66180×10$^{-21}$ | J=23←24 $K_a$=1←0 $K_c$=23←24 | N/A | Y |
| $^{32}SO_2$ | 430232.3126±0.0017 | 138.228 | 1.02560×10$^{-21}$ | J=20←21 $K_a$=1←2 $K_c$=19←20 | N/A | Y |
| $^{34}SO_2$ | 419070.9415±0.0056 | 222.079 | 5.31348×10$^{-22}$ | J=25←26 $K_a$=3←2 $K_c$=23←24 | Y | Y |
| $^{34}SO_2$ | 428537.9435±0.0065 | 167.768 | 1.63448×10$^{-21}$ | J=23←24 $K_a$=1←0 $K_c$=23←24 | Y | Y |
| $^{34}SO_2$ | 429952.4205±0.0075 | 188.460 | 8.54829×10$^{-22}$ | J=22 $K_a$=4←5 $K_c$=18←17 | Y | Y |
| $^{34}SO_2$ | 431498.3574±0.0061 | 179.871 | 1.07468×10$^{-21}$ | J=23←24 $K_a$=2←1 $K_c$=22←23 | Y | Y |
| $^{32}SO$ | 431808.196±0.020 | 67.731 | 8.78778×10$^{-21}$ | J=9←10 N=10←11 | Y | Y |
| $^{34}SO$ | 419640.353±0.014 | 68.080 | 6.61728×10$^{-21}$ | J=9←10 N=8←9 | Y | N |
| Na$^{35}$Cl | 428518.5512±0.0040 | 229.071 | 2.99626×10$^{-19}$ | J=32←33 | Y | Y |
| Na$^{37}$Cl | 419381.1264±0.0044 | 224.178 | 2.86536×10$^{-19}$ | J=32←33 | Y | Y |
| K$^{35}$Cl | 428297.8176±0.0027 | 393.948 | 1.58008×10$^{-19}$ | J=55←56 | Y | Y |
| K$^{37}$Cl | 416185.8003±0.0027 | 382.778 | 1.52012×10$^{-19}$ | J=55←56 | Y | N |
| K$^{37}$Cl | 430886.5999±0.0028 | 410.788 | 1.47360×10$^{-19}$ | J=57←58 | Y | N |
| Na$^{35}$Cl | 428518.5512±0.0040 | 229.071 | 2.99626×10$^{-19}$ | J=32←33 | Y | Y |
| Na$^{37}$Cl | 419381.1264±0.0044 | 224.178 | 2.86536×10$^{-19}$ | J=32←33 | Y | Y |

**Table S3. Best-fitting model parameters for local analysis.** Same as Table 1, but for the local regions shown in Fig S4, with 1σ uncertainties from MCMC simulations.

| Sulfur-bearing molecules | | | |
|---|---|---|---|
| Region | SO$_2$ column density [cm$^{-2}$] | T$_{gas}$ [K] | $^{34}$SO$_2$/$^{32}$SO$_2$ |
| Leading: EL | (1.306±0.034)×10$^{16}$ | 238.8±2.9 | 0.0579±0.0020 |
| Leading: WL | (7.19±0.34)×10$^{15}$ | 236.8±5.5 | 0.0574±0.0038 |
| Leading: DC | (1.971±0.090)×10$^{16}$ | 198.4±4.0 | 0.0526±0.0028 |
| Trailing: EL | (3.59±0.19)×10$^{15}$ | 260.5±7.9 | 0.0659±0.0051 |
| Trailing: WL | (4.22±0.31)×10$^{15}$ | 240.3±9.4 | 0.0660±0.0061 |
| Trailing: DC | (5.61±0.55)×10$^{15}$ | 210.2±9.7 | 0.0682±0.0081 |

| Chlorine-bearing molecules | | | | | |
|---|---|---|---|---|---|
| Region | NaCl column density [cm$^{-2}$] | KCl column density [cm$^{-2}$] | T$_{gas}$ [K] | $^{37}$Cl/$^{35}$Cl | Velocity [m s$^{-1}$] |
| Leading: 1 | (4.18±0.23)×10$^{12}$ | (7.79±0.77)×10$^{11}$ | 934±59 | 0.405±0.020 | 19±11 |
| Leading: 2 | (4.99±0.23)×10$^{12}$ | (5.3±1.4)×10$^{11}$ | 727±49 | 0.348±0.021 | 52±11 |
| Leading: 3 | (2.82±0.40)×10$^{11}$ | (4.0±1.1)×10$^{10}$ | 1320±180 | 0.399±0.044 | 275±22 |
| Trailing: 1 | (4.38±0.13)×10$^{12}$ | (5.4±1.1)×10$^{11}$ | 639±28 | 0.419±0.018 | -90.3±6.4 |
| Trailing: 2 | (7.12±0.31)×10$^{11}$ | (7.5±1.3)×10$^{10}$ | 796±42 | 0.417±0.016 | -150±11 |
| Trailing: 3 | (6.21±0.62)×10$^{11}$ | (3.7±2.1)×10$^{10}$ | 1290±130 | 0.350±0.029 | -24±15 |

**Table S4. NaCl and KCl source locations and possible identifications with paterae.** Location numbers correspond to Figure 1. The most likely patera is identified if there is a clear, isolated patera at the identified latitude and longitude. All other named paterae within the uncertainties on the latitudes and longitudes are also listed. The latitudes and longitudes of leading hemisphere location 1 and trailing hemisphere location 3 are consistent within 1σ of one another so might be the same gas source.

| Location | Latitude | Longitude | Most likely patera | Other paterae within uncertainties |
|---|---|---|---|---|
| Leading hemisphere | | | | |
| 1 | 27±10ºS | 20±14ºW | | Kanehekili Fluctus, Cataquil Patera, Uta Patera, Angpetu Patera |
| 2 | 66±11ºS | 136±17ºW | | |
| 3 | 22±10ºN | 159±14ºW | | Thomagata Patera, Reshef Patera, Surya Patera, Chaac Patera |
| Trailing hemisphere | | | | |
| 1 | 58±10ºS | 218±17ºW | Kurdalagon Patera | Gabija Patera |
| 2 | 30±11ºN | 211±13ºW | Isum Patera | Susanoo Patera |
| 3 | 24±10ºS | 355±15ºW | | many |
| 4 | 33±12ºN | 330±15ºW | Fuchi Patera | Manua Patera |
| 5 | 59±11ºS | 341±15ºW | Creidne Patera | Hiruko Patera, Inti Patera |

**Table S5.** Data sources for previous isotope measurements shown in Fig 4.

|  | Sample | Reference |
|---|---|---|
| **Sulfur** | Solar wind | (*62*) |
|  | Bulk Silicate Earth (BSE) | (*63*) |
|  | Earth sediments | (*64*) |
|  | Earth volcanic gasses | (*65*) |
|  | Lunar melt inclusions (MIs) | (*66*) |
|  | Lunar mare basalts | (*67*) |
|  | Gale crater sediments | (*68*) |
|  | Ordinary chondrites (OC) | (*28*) |
|  | Comet Hale-Bopp $H_2S$ | (*69*) |
|  | Comet Hale-Bopp CS | (*70*) |
|  | Comet 67P/Churyumov-Gerasimenko | (*29*) |
|  | Comet C/2014 Q2 (Lovejoy) | (*71*) |
|  | Comet C/2012 F6 (Lemmon) | (*71*) |
|  | Interstellar medium (ISM) | (*72*) |
|  | Galactic cosmic rays | (*73*) |
| **Chlorine** | Venus, HCl gas | (*74*) |
|  | Moon, basalts and soils | (*75, 76*) |
|  | Mars, HCl gas | (*77*) |
|  | Vesta, from apatite in eucrites | (*78*) |
|  | Comet 67P/Churyumov-Gerasimenko, from HCl gas | (*79*) |